\documentclass[prb,twocolumn,showpacs,amsmath,amssymb,superscriptaddress]{revtex4-1}

\usepackage{graphicx}
\usepackage{dcolumn}
\newcolumntype{d}{D{.}{.}{4.1}} 
\usepackage{bm}
\usepackage{multirow}
\usepackage{color}
\usepackage{hyperref}
\usepackage{float}
\hypersetup{pdfnewwindow=true, colorlinks=true, linkcolor=blue, anchorcolor=blue, citecolor=blue, filecolor=blue, menucolor=blue, urlcolor=blue}

\begin{document}

\title{Hydrogen plasma favored modification of anatase TiO$_2$ (001) surface with desirable water splitting performance}

\author{Ming Lei}
\email{mlei012@ucr.edu}
\affiliation{Chemical and Environmental Engineering, University of California Riverside, CA 92521, USA}
\author{Sinisa Coh}
\affiliation{Materials Science and Mechanical Engineering, University of California Riverside, CA 92521, USA}

\date{\today}

\begin{abstract}
We show that when TiO$_2$ anatase (001) is exposed to hydrogen plasma that the pristine surface termination becomes unfavorable to another, slightly modified, surface. On this modified surface the topmost TiO$_2$ layer is intact but out of registry with the bottom layers. Nevertheless, the modified surface has significantly improved ability to split water under exposure to sunlight. We show by explicit calculation of the water splitting reaction that the energy barrier that exists on a pristine surface is not present on the modified surface.  The valence band maximum of the surface is raised relative to the pristine surface, which is a favorable way of adjusting the band gap in TiO$_2$ to the solar spectrum.
\end{abstract}

\maketitle

\section{Introduction}

Anatase TiO$_{2}$ is a promising and well-studied material for photocatalytic water splitting.  TiO$_2$ is cheap, environment-friendly, and stable. As is well known, due to its large band gap, TiO$_2$ can't utilize the sunlight efficiently. Furthermore, to provide enough driving force for the photo-oxidation of water, the valence band maximum (VBM) of TiO$_2$ should be increased and positioned closer to, but lower than, the potential of H$_2$O/O$_{2}$ pair.\cite{doi:10.1021/jp711929d} Up to now, plenty of research has been conducted to narrow its band gap and to elevate the VBM.

Doping is one common strategy to reduce the band gap of TiO$_{2}$. Doping can be done with either a metal ion (La,\cite{LIQIANG20043375} V,\cite{doi:10.1021/jp004295e} Co,\cite{doi:10.1021/ie061491k} Nb,\cite{doi:10.1002/adfm.200901292} Ag\cite{SOBANA2006124}) or a nonmetal ion (S,\cite{doi:10.1246/cl.2003.364} B,\cite{doi:10.1021/ja0749237} C,\cite{doi:10.1246/cl.2003.772,doi:10.1021/nl051807y} N\cite{doi:10.1021/jp051756t}). However, some dopants, especially the metal ions, can lead to severe carrier recombination which reduces the overall quantum efficiency.\cite{doi:10.1021/j100102a038}  Moreover, the metal ion dopants are often polluting.

An alternative approach for narrowing the band gap, and raising the VBM, is to modify the surface structure of anatase TiO$_{2}$ by hydrogenation.\cite{Chen746}  This treatment changes the color of anatase TiO$_{2}$ from white to black,\cite{C4CS00330F,Chen746,PhysRevLett.111.065505} and produces highly rough and amorphous surface of TiO$_{2}$, both in the case of nanoparticles \cite{TENG2014339,C3TA13491A} and nanotubes.\cite{doi:10.1002/adfm.201303042,wu_2013} Furthermore, according to the x-ray photoelectron spectroscopy (XPS) measurement, the band gap narrowing of black TiO$_2$ is achieved by raising the VBM without changing the CBM (conduction band minimum)\cite{Chen746} which is a favorable band alignment for photocatalytic water splitting. While it remains unknown exactly which structural modification leads to the favorable band alignment of black TiO$_2$, there is evidence that likely the increase in VBM is due to the surface modification of TiO$_2$.\cite{Chen746,PhysRevLett.111.065505} On the other hand, XPS study on a similar system (nanowires instead of nanoparticles) by Wang {\it et al.}\cite{doi:10.1021/nl201766h} found no shift in the valence band.  They assigned the dark color of nanowires to the formation of defect states or impurities.  Furthermore, Alberto and co-workers\cite{doi:10.1021/ja3012676} synthesized black TiO$_{2}$ nanoparticles with crystalline core and disordered shell morphology and found that in addition to the surface modification, the presence of oxygen vacancies could also contribute to the visible light absorption of TiO$_2$. 

In this work, we explore possible structural changes to the (001) surface of TiO$_2$ after the adsorption of hydrogen atoms. Our calculations show that the structure of pristine (001) surface of TiO$_2$ anatase becomes unfavorable to another structure with a slight modification on the surface when exposed to a hydrogen atom pressure of 0.8--80~Pa, which can be achieved in the laboratory.\cite{NAKAMURA2000205}  Furthermore, by doing an explicit calculation of the water splitting process, we find that the structural modification removes the rate-limiting step for the water splitting process. More specifically, while the pristine anatase TiO$_{2}$ (001) surface has a barrier of 0.44~eV in the Gibbs free energy profile of the oxygen evolution reaction, there is no such barrier on the modified surface.  Therefore, our calculations show that the hydrogen-plasma treated anatase TiO$_{2}$ (001) surface is a suitable candidate for water splitting applications. 

While in this work we focus on the (001) surface, it is possible that similar modifications exist on other TiO$_2$ surfaces, such as majority (101) surface. However, the structural modification we studied consists of Ti atom displacements in the plane parallel to the (001) surface,\cite{PhysRevB.95.085422} so it is the most natural to study the modification of the (001) surface.  Finally, we note that while the (101) surface is the majority surface in anatase TiO$_2$ it is not very reactive, and the (001) surface plays an important role in reaction.\cite{doi:10.1021/jp0275544,PhysRevLett.81.2954,doi:10.1021/jp055311g}

This paper is organized as follows: In Sec.~\ref{sec:details} we describe the calculation details. In Sec.~\ref{sec:surface} and Sec.~\ref{sec:water} we present and discuss our results. We give an outlook and conclude in Sec.~\ref{sec:conclusion}.

\section{Calculation details}
\label{sec:details}

For our calculations we use density functional theory as implemented in the Quantum Espresso package.\cite{Giannozzi_2009} We use the generalized gradient approximation (GGA) of Perdew, Burke, and Ernzerhof (PBE)\cite{PhysRevLett.77.3865} along with the ultrasoft pseudopotentials from the GBRV database.\cite{GARRITY2014446} These pseudopotentials describe the valence electrons 3{\it s}3{\it p}3{\it d}4{\it s} in Ti, 2{\it s}2{\it p} in O and 1{\it s} in H. Since PBE functional fails to provide the correct description of a localized Ti$^{3+}$ state as a result of H adsorption, we adopt GGA+{\it U} approach, and we use the {\it U} value of 4~eV on the Ti 3{\it d} states following Ref.~\onlinecite{ggau}. We also include spin polarization in the calculation. In order to obtain sufficient precision, we cutoff the plane wave basis for the wavefunction at 40~Ry and 400~Ry for the density. We use 15~\AA\ of vacuum to avoid the interaction between neighboring slabs. In cases when the surfaces of the slab are different, for example when molecules are absorbed on one of the sides of the slab, we use the dipole correction in the direction perpendicular to the surface.  All surface energies in the paper are reported per one side of the slab.  We sample the electron's Brillouin zone on a 6$\times$6$\times$1 Monkhorst-Pack grid.

To model a slab of TiO$_2$ we set the in-plane lattice constant of the slab equal to the in-plane bulk lattice constant. Each time we fully relax the slab with the only constraint that the in-plane lattice constant remains unchanged. We use a slab with a thickness of eight layers of TiO$_{2}$. For most calculations we use minimal in-plane unit cell, but for the oxygen evolution reaction processes, we use an in-plane supercell that is doubled along with one of the in-plane lattice vectors.

We calculate the surface energy density $\Delta \gamma$ as 
\begin{align}
\Delta \gamma & = \frac{1}{A} \left[ E_{\rm slab}-E_{\rm clean}-N_{\rm H} \mu_{\rm H}(T,p) \right]
\end{align}
Here $E_{\rm slab}$ is the energy of the slab with $N_{\rm H}$ hydrogen atoms adsorbed on the surface. $E_{\rm clean}$ is the energy of the slab without adsorbed hydrogen atoms. The surface area is $A$. Following Ref.~\onlinecite{PhysRevB.65.035406}, $\mu_{\rm H}(T,p)$ is the chemical potential of H atom as a function of temperature ($T$) and pressure ($p$),
\begin{align}
    \mu_{\rm H} (T,p^0) & = {\cal H}_{\rm H} (T,p^0)-{\cal H}_{\rm H}(0,p^0) \notag \\
    & - T[{\cal S}_{\rm H}(T,p^0)-{\cal S}_{\rm H}(0,p^0)] \label{mu} \\
\mu_{\rm H}(T,p) & = \mu_{\rm H} (T,p^0)+ k_{\rm B} T \ln (p/p^0) \label{delta}.
\end{align}
Here, $p^0$ is the pressure of a reference state, and $\mu_{\rm H} (T,p^0)$ is the chemical potential of H atom at the reference state with temperature $T$ and pressure $p^0$. The temperature dependence of the enthalpy $\cal H$ and entropy $\cal S$ at the reference state are tabulated in the thermochemical reference tables.\cite{stull1971janaf} We chose $T=700$~K in our calculations.

\section{Hydrogen plasma favored surface modification}
\label{sec:surface}

Now we present our results on the structural modification of the TiO$_2$ anatase $(001)$ surface under the hydrogen plasma environment. In the next section we will study its water splitting performance.

\begin{figure}
\includegraphics[width=3.4in]{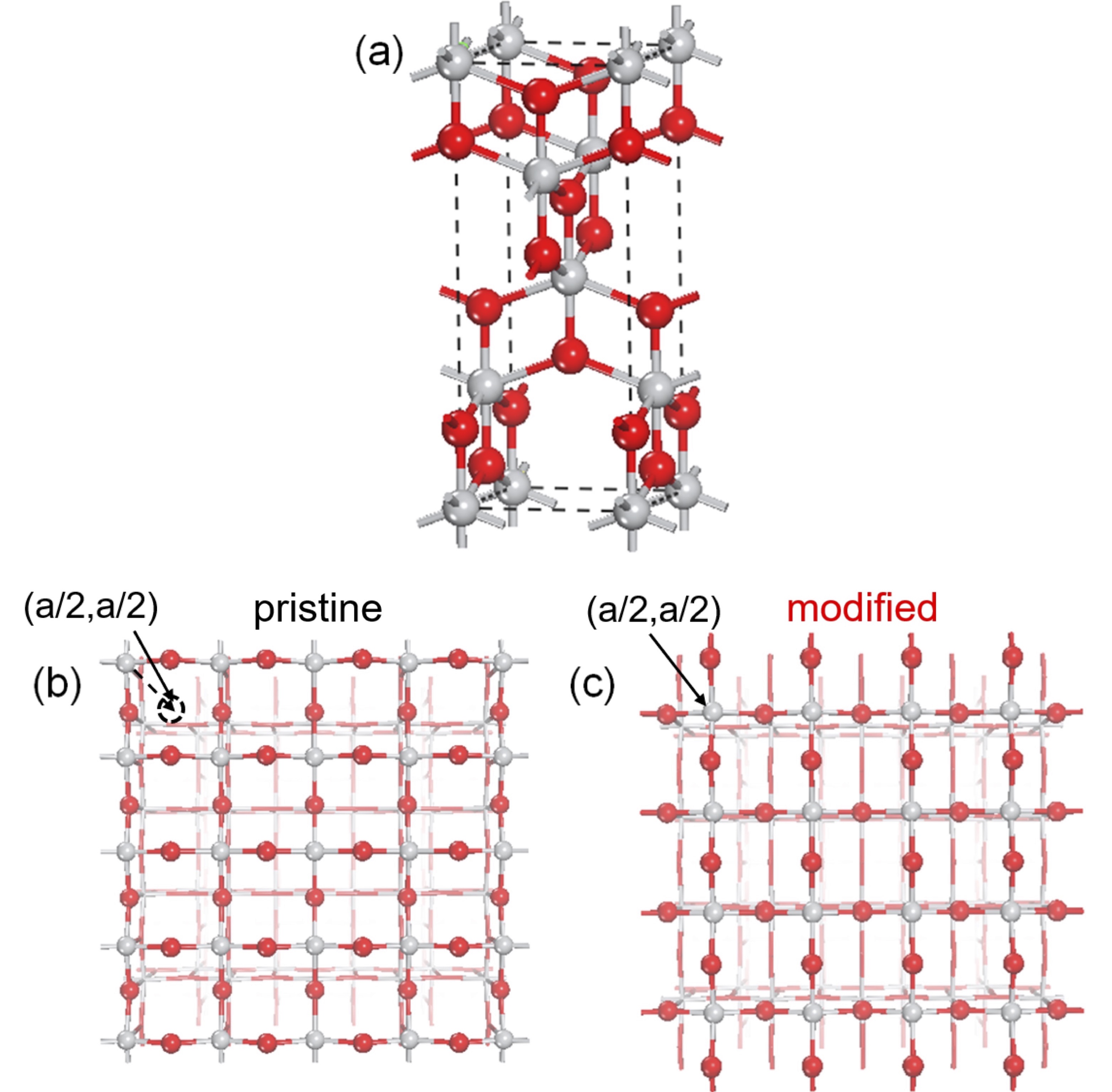}
\caption{\label{fig:bulk}The conventional unit cell of bulk anatase TiO$_{2}$ (a).  The pristine (b) and modified (c) surfaces of anatase TiO$_2$ differ only by translation of the top-most layer of TiO$_2$ by vector $(a/2,a/2)$. Ti, gray; O, red.}
\end{figure}

The conventional unit cell of bulk anatase TiO$_{2}$ is shown in Fig.~\ref{fig:bulk}a.  Anatase TiO$_2$ crystallizes in space group {\it I} $4_1/${\it amd} (space group 141).  Titanium atoms are at  Wyckoff orbit $b$ while oxygen atoms are at Wyckoff orbit $e$.  Therefore, all titanium atoms in the crystal structure are equivalent to each other, and all oxygen atoms are equivalent to each other.  Our calculated relaxed lattice parameters of bulk anatase TiO$_{2}$ are $a$ = $b$ = 3.804~\AA\ and $c$ = 9.695~\AA, which is close to the experimental result $a$ = $b$ = 3.804~\AA\ and $c$ = 9.614~\AA\ (0.84\% deviation).\cite{horn1972}  The structural unit of TiO$_2$ anatase is a TiO$_6$ octahedron with Ti atom in the center of the octahedron, and O atoms in the corners of the octahedron.  These octahedra are connected to each other and are forming an edge-sharing network.  Each O atom is bonded to three Ti atoms.

Based on our calculation of bulk anatase TiO$_{2}$ we constructed a model of pristine $(001)$ surface.  This surface is shown in Fig.~\ref{fig:bulk}b. While in the bulk TiO$_2$ all Ti atoms are six fold coordinated, this is clearly not the case on the $(001)$ surface.  Here, the breaking of the Ti--O bond perpendicular to the surface reduces the coordination of the topmost Ti atom from six to five. Furthermore, there are now two symmetry inequivalent oxygen atoms at the surface: only one of which is nominally saturated, as it is surrounded by three Ti atoms.  Another oxygen atom is unsaturated, as it is surrounded by only two Ti atoms.  Therefore, all Ti and half of the O atoms at the topmost layer of the TiO$_2$ (001) surface are nominally unsaturated.

\subsection{Hydrogen adsorbed on a pristine surface}

\begin{figure}
\includegraphics[width=3.4in]{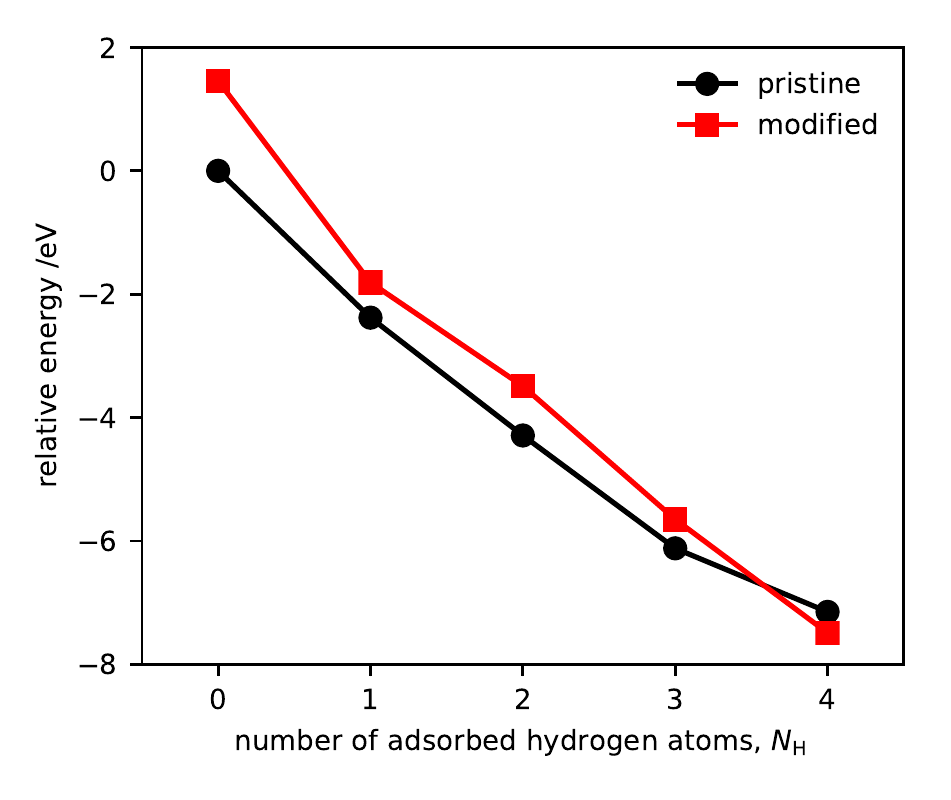}
\caption{\label{fig:stability}Relative stability of pristine and modified surfaces after adsorption of hydrogen atoms.}
\end{figure}

We now discuss the hydrogen passivization of the (001) surface. We consider all the Ti and O atoms on both the first and second layers of the surface as potential hydrogen atom adsorption sites. Relative energies of sites with strongest (optimal) absorption are shown in Fig.~\ref{fig:stability} as a function of number of adsorbed H atoms.

We note that hydrogen atoms could also diffuse deeper into the bulk, beyond the first and second layers we considered here. As discussed in Refs.~\onlinecite{doi:10.1021/jp200408v,doi:10.1021/jp210472p} the barriers for diffusion into the bulk and onto the surface of TiO$_2$ are similar.  Diffusion into the bulk could additionally change the surface chemistry via strain. Furthermore, the induced strain could also change the relative energy of the pristine and modified surface of TiO$_2$.  However, the main goal of this work is to isolate the role of surface modification on surface chemistry, so we don't discuss further the role of diffusion into the bulk.

When only one H atom is adsorbed per surface unit cell we find that the H atom prefers to absorb horizontally on the surface two-coordinated O atom. The adsorption energy is $-2.38$~eV. The negative sign for the adsorption energy means that it is energetically favorable for H atom to absorb on the surface. This binding energy is specified relative to a single isolated H atom.  With two adsorbed H atoms, one prefers to adsorb again to the topmost two-coordinated O atom, while the second adsorbs on the surface Ti atom.  The total adsorption energy for these two H atoms taken together is $-4.29$~eV.  Therefore, adsorption energy per atom is now decreased from $-2.38$~eV to $-4.29/2=-2.15$~eV.  With three adsorbed H atoms we find that first two adsorb as before, while the third one prefers to adsorb on the O atom in the second layer.  In this case the total adsorption energy for all three atoms is $-6.12$~eV.  Finally, we find that with four adsorbed H atoms, two are adsorbed on the first layer and another two on the second layer. The total adsorption energy for these four H atoms is $-7.15$~eV.  The adsorption energy per hydrogen atom in this case is therefore reduced to $-7.15/4=-1.79$~eV.

We find that the pristine (001) surface can't absorb more than four H atoms per cell.  If we try adding the fifth H atom to the surface, we find that the added H atom combines with another H atom on the surface to generate a H$_2$ molecule and moves away from the surface. Therefore, we conclude that the unsaturated pristine surface of (001) TiO$_2$ anatase can be saturated by adsorption of at most four H atoms per surface unit cell.

In all of these cases we fully relaxed the TiO$_2$ surface in the presence of hydrogen atoms.  The surface relaxation is significant, with the maximal atomic displacements on the order of 0.4~\AA.   As one would expect, surface relaxation is smallest with one adsorbed hydrogen atom and largest with four adsorbed hydrogen atoms.  

Without inclusion of +{\it U} correction on Ti we find that the absorption energies are slightly different. At the GGA level we find that H-adsorption energies are $-2.23$~eV, $-4.06$~eV, $-5.33$~eV, and $-6.85$~eV, as compared to $-2.38$~eV, $-4.29$~eV, $-6.12$~eV, and $-7.15$~eV at the GGA+{\it U} level.

\subsection{Hydrogen adsorbed on a modified surface}

Now we consider the possibility that the presence of adsorbed hydrogen atoms will lead the (001) surface to reconstruct in a distinct basin of energy, with a structure that is different from the pristine (001) surface.

Indeed, our calculations show that when four hydrogen atoms are adsorbed per a single surface unit cell that the surface of TiO$_2$ (001) anatase is reconstructed.  This reconstruction is similar to the bulk structure discussed in Ref.~\onlinecite{PhysRevB.95.085422}.  The structure can be approximately described as the translation of Ti atoms on the surface by $(a/2,a/2)$ relative to the pristine surface.  This translation vector is shown in Fig.~\ref{fig:bulk}. (Alternatively, this structure could be obtained by rotating the topmost layer by 90$^{\circ}$ around either of the surface oxygen atoms.)

Note that if the topmost layer was isolated from the other layers, that the pristine and reconstructed surface would be indistinguishable from each other.  This can easily be seen from Fig.~\ref{fig:bulk} or by realizing that in a surface primitive unit cell oxygen atoms are at coordinates $(0,a/2)$ and $(a/2,0)$ while titanium atom is at the origin $(0,0)$.  Therefore, if we translate titanium atom by $(a/2,a/2)$ we get a crystal structure equivalent to the one where we translate all atoms by $(a/2,a/2)$, as oxygen atoms get mapped into periodic images of each other.  Therefore, this modified structure is in some sense minimally perturbed relative to the pristine surface, as the only difference of the modified surface is that the topmost layer is out of registry with the rest.

We studied the adsorption of hydrogen atoms to the modified surface following steps analogous to those used for the pristine surface.  The calculated relative energy of modified surface adsorbed with a different number of H atoms is indicated in Fig.~\ref{fig:stability} with red color.  We took the energy of the pristine surface without H atom adsorption as a reference state with zero energy.

We find that without H atom adsorption, the energy of the modified surface is 1.46~eV higher than that of the pristine surface.  However, once H atoms are adsorbed the energy difference between modified and pristine surface diminishes. Eventually, with four adsorbed H atoms the modified surface becomes energetically favorable compared to the pristine surface.

More specifically, we find that the absorption energy of the first hydrogen atom is $-3.27$~eV.  While this surface absorbs hydrogen atom slightly more strongly than the pristine surface ($-3.27$~eV compared to $-2.38$~eV on a pristine surface), the difference is not large enough to compensate for the increased surface energy of the modified surface relative to the pristine surface (1.46~eV).  However, if we increase the number of hydrogen atoms to two per surface unit cell, the total adsorption energy increases to $-4.95$~eV (compared to $-4.29$~eV in the pristine case).  With three hydrogen atoms, it is $-7.11$~eV, and finally, it is $-8.95$~eV with four H atoms. Therefore, when four H atoms are adsorbed, the modified surface becomes favorable relative to the pristine surface, as the stronger preference of H adsorption on the modified surface ($-8.95$~eV versus $-7.15$~eV gives the energy difference $-7.15-(-8.95)=1.80$~eV) is large enough to compensate for the difference in the surface energy between the modified and pristine surface (1.46~eV). However, the energy difference between pristine and modified surfaces with four adsorbed H atoms is only moderate ($1.80-1.46=0.34$~eV), and it might be comparable to the error of the approximations used in our calculation. Nevertheless, our calculation clearly shows that the modified surface has a tendency to absorb more hydrogen atoms and will thus be energetically more and more favorable at high enough pressure of hydrogen atoms.

We again find that without the inclusion of +{\it U} correction on Ti that the absorption energies are somewhat different. For example, when four H atoms are adsorbed, we find that the modified surface is preferred over the pristine surface by $(1.80-1.46)$~eV at the GGA+{\it U} level and $(1.49-1.40)$~eV at the GGA level.  Therefore, including +{\it U} correction changes the adsorption energy of H atoms and it further increases the stability of the modified surface relative to the pristine surface.

\subsection{Required hydrogen pressure}

\begin{figure}
\includegraphics[width=3.4in]{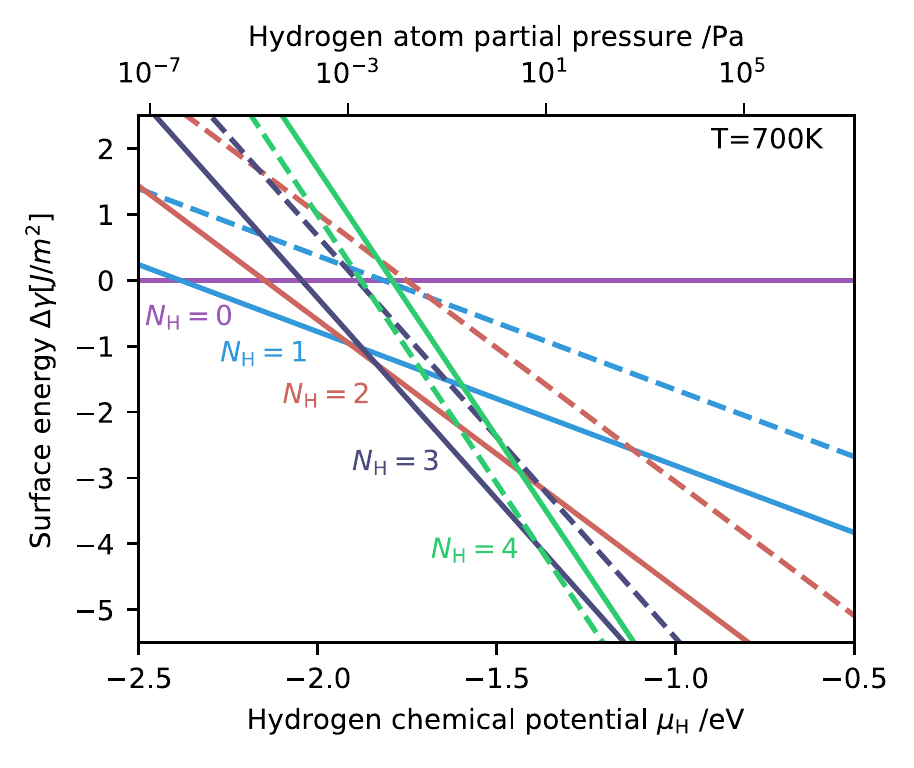}
\caption{\label{fig:potential}The surface energy density $\Delta \gamma $ of the pristine (solid) and modified (dashed) surfaces with $N_{\rm H}=0$ through 4.  Above $\mu_{\rm H}=-1.38$~eV modified surface (dashed green) becomes the most favorable surface.}
\end{figure}

Based on the surface adsorption energies of H atoms we will now determine the required pressure of hydrogen atoms needed to modify the pristine TiO$_2$ surface.  Figure~\ref{fig:potential} shows the surface energy $\Delta \gamma$ of the pristine and modified surfaces as a function of the hydrogen chemical potential $\rm \mu_H$.  As can be seen from the figure, surface without hydrogen atom adsorption $N_{\rm H}=0$ is favorable at hydrogen chemical potential below $-2.38$~eV.  As hydrogen chemical potential is increased, the preferred surface becomes the pristine surface with $N_{\rm H}=1$ hydrogen atom adsorption.  At $\mu_{\rm H} = -1.91$~eV the preferred surface becomes the pristine surface with $N_{\rm H}=2$.  Finally, when the hydrogen chemical potential is larger than $-1.38$~eV, the modified surface with $N_{\rm H}=4$ becomes the most favorable surface. Since a small difference in the chemical potential can cause a large change of the pressure, based on equations (\ref{mu}) and (\ref{delta}), we report the needed pressure of hydrogen atoms as a range 0.8--80~Pa.  This range assumes a 10\% error in the calculated hydrogen chemical potential.

\subsection{Energy barrier between pristine and modified surface} 

\begin{figure}
\includegraphics[width=3.4in]{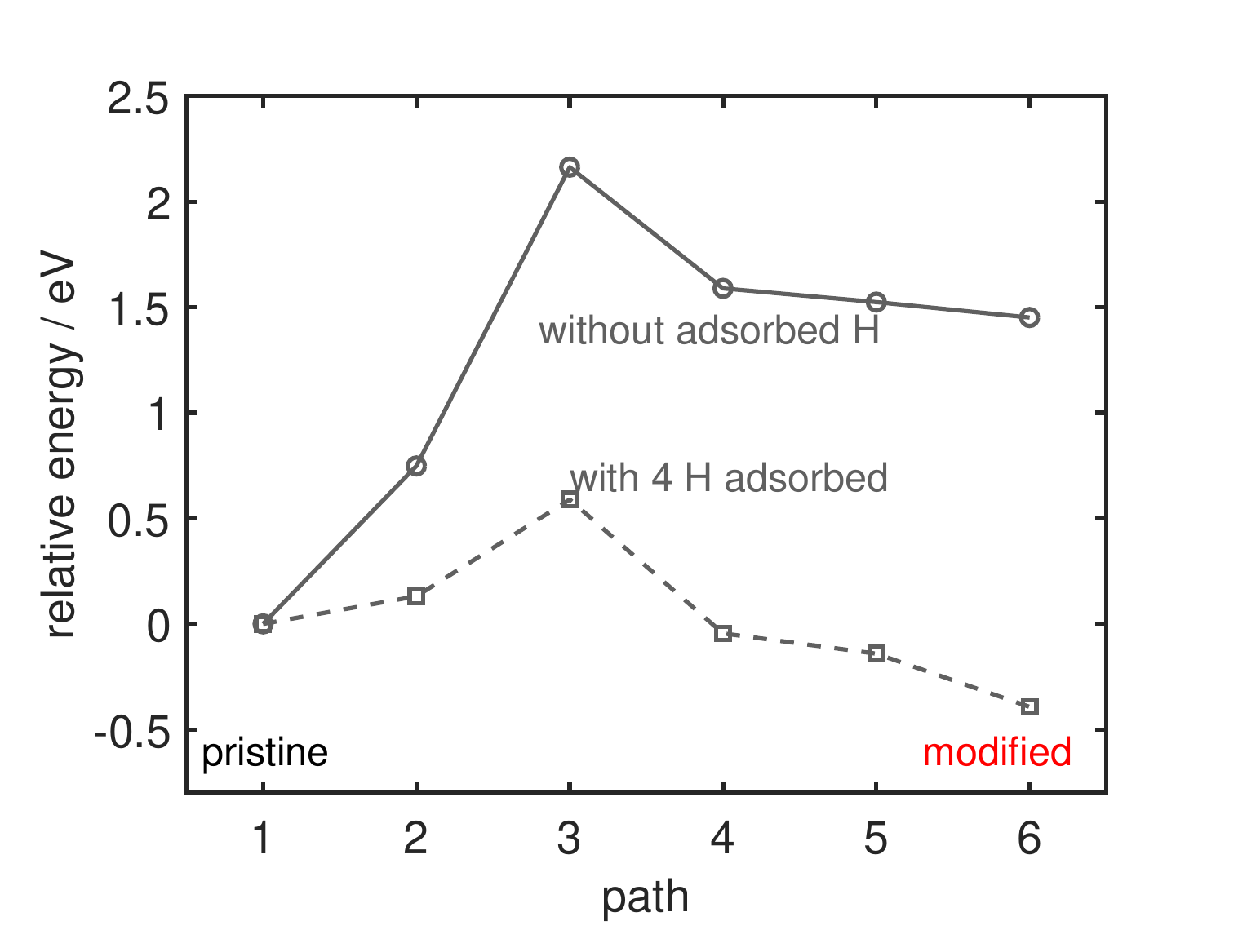}
\caption{\label{fig:barrier}Calculated energy profiles for the structure transformation from the pristine to the modified structure, both without (solid line) and with (dashed line) adsorbed hydrogen.}
\end{figure}

Now we turn to calculate the barrier between the pristine and modified surface.  We calculated the barrier using the nudged elastic band (NEB) approach.  As shown in Fig.~\ref{fig:barrier} the barrier is very high when there are no hydrogen atoms adsorbed on the surfaces (it is 2.16~eV, per primitive surface unit cell).  However, once four hydrogen atoms are adsorbed on the surface, the barrier is reduced to only 0.59~eV per primitive surface unit cell.  The reduced energy pathway between pristine and modified surface is indicated with a dashed line in Fig.~\ref{fig:barrier}.  Furthermore, from the NEB calculation we infer that the modified structure remains metastable even when the hydrogen atoms are removed, so that even without hydrogen chemical potential there is a barrier for the modified surface to revert to the pristine surface.

\subsection{Other potential structural modifications} 

So far we only discussed one reconstruction of TiO$_2$ anatase (001) surface in the presence of hydrogen atoms.  In what follows we consider some other possible reconstructions compatible with the minimal 1$\times$1 surface unit cell and show that they are all energetically less stable.  It is possible that a significantly larger computational unit cell could lead to different surface reconstruction.  However, we leave this to future studies, as with increased unit cell size the calculations quickly become very computationally demanding.  Nevertheless, we confirmed that doubling the unit cell size gives the same trend for the binding energy of H to the pristine and modified surface.

The modified structure of TiO$_2$ we discussed earlier can be constructed by translating the topmost Ti atom by $(a/2,a/2)$.  This translation breaks one of the Ti-O bonds, the one that is perpendicular to the surface, which is why the energy of the surface increases by 1.46~eV.  However, as discussed earlier, this energy difference is compensated by the fact that the surface with a broken Ti-O bond can absorb more hydrogen atoms.  Motivated by this finding, we will now consider different ways to break Ti-O bonds on the surface and check whether they can also be compensated energetically by absorbing additional hydrogen atoms.

The first alternative way to break the Ti-O bond we considered was to simply increase the vertical distance between Ti and O atoms.  If we try inserting an additional H atom between the bond-breaking Ti and O, we find that instead of the formation of Ti-H or O-H bond, hydrogen atoms bind together and form a H$_2$ molecule inside the slab. The energy of this structure is 0.27~eV higher than the total energy of the pristine surface adsorbed with four H atoms (plus one isolated H atom, to keep the total number of H atoms the same). Therefore, we conclude that hydrogen atoms can't stabilize the breaking of the vertical Ti-O bond, unless one translates Ti atom by $(a/2,a/2)$, as in the modified structure. 

The second structure we tried has a broken Ti-O bond that is parallel to the surface. As in the previous case, we broke the bond simply by increasing the distance between the Ti and O atoms in the bond. If we try adding two additional H atoms between Ti and O, we find that two H$_2$ molecules are generated during the structural relaxation. The energy of this configuration is 0.31~eV higher than the total energy of the pristine surface adsorbed with four H atoms (plus the energy of two isolated H atoms). Therefore, we conclude that in this scenario hydrogen atoms can't break the in-plane Ti-O bond. 

In the third structure, we tried to contain oxygen vacancy at the surface. To form a vacancy we removed one surface two-coordinated O atom.  Next we tried to stabilize this surface by putting one additional H atom at the vacancy site. The formation energy of this oxygen vacancy is 1.64~eV, so this structure can't be stabilized by the addition of hydrogen atom.  We quantified the formation energy of the vacancy by taking a difference between the total energy of the slab with a vacancy (plus half isolated O$_2$ molecule) and the total energy of the pristine surface adsorbed with four H atoms (plus one isolated H atom).

\begin{figure}
\includegraphics[width=3.4in]{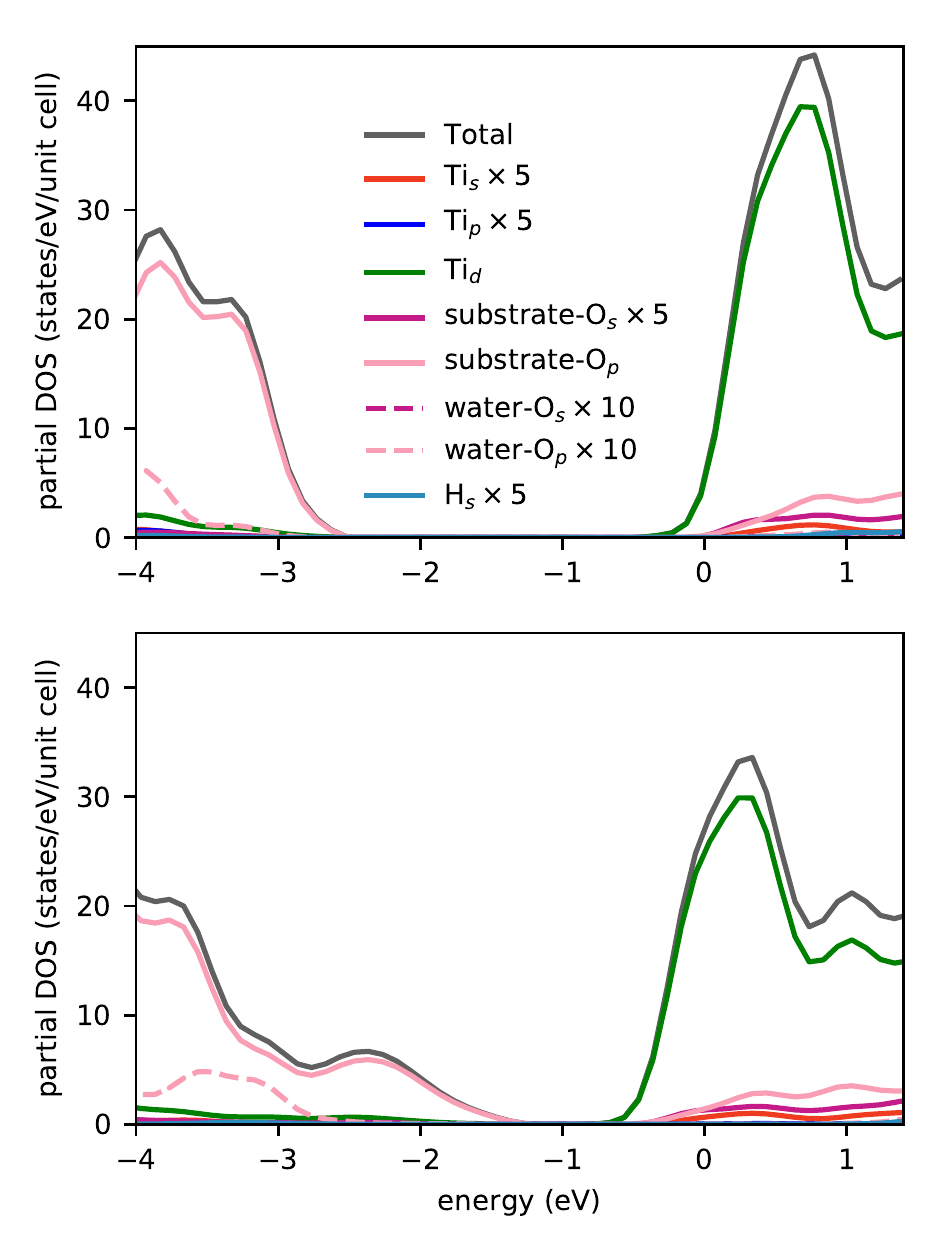}
\caption{\label{fig:doswater} The projected density of states of Ti, O, and H orbitals for pristine (up) and modified (bottom) surface of TiO$_2$.  In each case one water molecule is physically adsorbed on the surface and no hydrogen atoms are adsorbed to the surface.  The amplitudes of Ti$_s$, Ti$_p$, substrate-O$_s$, and H$_s$ projected densities of states are multiplied by 5, and the heights of water-O$_s$ and water-O$_p$ are multiplied by 10.}
\end{figure}

\section{Water splitting performance of the modified surface}
\label{sec:water}

After showing that the modified surface becomes favorable at high hydrogen plasma pressure, now we turn to the study of the water splitting performance of the modified surface.  In what follows we study the modified surface on its own, without the presence of the adsorbed hydrogen atoms.  As shown in Fig.~\ref{fig:barrier} the modified structure remains metastable even when adsorbed hydrogen atoms are removed from the surface.

\begin{figure*}
\includegraphics{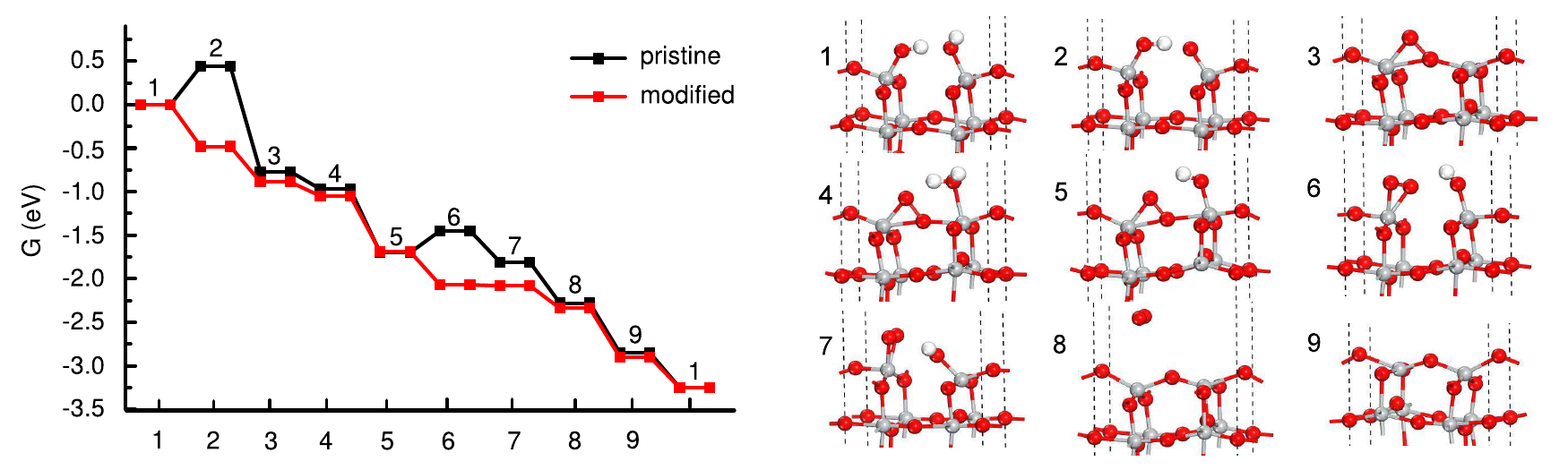}
\caption{\label{fig:path}(left) Gibbs free energy profiles of oxygen evolution reaction on pristine (001) and modified (001) surface (1.93~V versus the standard hydrogen electrode). (right) Optimized structures of the intermediate states on the modified (001) surface. Ti, gray; O, red; H, white. }
\end{figure*}

\subsection{Water adsorption}

When anatase TiO$_2$ is used to catalyze water splitting, the catalyst is immersed in the aqueous environment. Therefore, we will first compare the adsorption of water molecule on pristine and modified surfaces.  We find that after full structural relaxation the water molecule binds to the pristine surface by forming a bond between the Ti atom on the surface and O atom of the water molecule (bond length is 2.33~\AA).  One of the H atoms in the water molecule interacts with the surface O atom and generates a bond with length of 1.58~\AA.  We expect that a very similar binding geometry will occur on the modified surface, as the top layer is nearly the same as in the pristine case.  This is precisely what we find, as the Ti-O bond length on the modified surface is 2.30~\AA\ while the H-O bond length is 1.59~\AA. 

Despite similarities in the structure, the energy level alignments are not the same on two surfaces.  Figure~\ref{fig:doswater} shows the calculated density of states (DOS) of the pristine and modified surfaces with adsorbed water molecule. The energy in that figure is aligned relative to the vacuum level above the surfaces. With GGA+{\it U} calculation, the band gap of the pristine surface is about 2.8~eV, which is larger than the band gap calculated by GGA (1.8~eV)\cite{doi:10.1021/acs.jpcc.7b02964} and closer to the GW band gap (3.5~eV).\cite{PhysRevB.89.075205} Our electronic structure calculations are consistent with experimental findings from Ref.~\onlinecite{Chen746} on the so-called black TiO$_2$. Both experiment Ref.~\onlinecite{Chen746} and our calculation find that the valence band maximum of the modified surface is increased in energy relative to the pristine surface, while the conduction band minimum is nearly unchanged. Furthermore, as indicated in the figure, the shift of the valence band maximum originates from the oxygen 2{\it p} states in TiO$_2$ and not from the oxygen 2{\it p} states in the water molecule.

\subsection{Reaction path}

Since the modified surface has a higher VBM than the pristine surface, the reduced band gap of the modified surface should contribute to better water splitting performance.  To test this hypothesis, we studied the explicit chemical reaction on the surfaces. Several groups have already studied theoretically the mechanism of photocatalytic water splitting on the pristine anatase TiO$_2$ surface.\cite{doi:10.1021/ja105340b,doi:10.1021/jp212407s}  Since pristine and modified surface differ only in the registry between the two top-most layers of TiO$_2$, we suspect that the reaction pathways might be similar.  After all, the topmost layer of TiO$_2$ that is exposed to the water molecules is structurally nearly the same in two cases. However, the magnitude of energy barriers needs not be the same, as electronic structures are different. 

As a reference path for the pristine TiO$_2$ surface we took the reaction path proposed by Liu {\it et al.} in Ref.~\onlinecite{doi:10.1021/ja105340b}.  We didn't implement the solvation effects in our calculations as solvation effects have a small effect on the relative energy of each state.  For example, we find that the rate-limiting barrier without solvation effects is 0.44~eV while Ref.~\onlinecite{doi:10.1021/ja105340b} reports that the same barrier is 0.61~eV with solvation effects.

Figure~\ref{fig:path} shows calculated Gibbs free energy profiles of oxygen evolution reaction on the pristine (black line) and modified surface (red line), as well as the optimized structures of the intermediate states on the modified (001) surface. State 1 on the figure represents the surface adsorbed with one dissociated water molecule where one H$^+$ ion is adsorbed on the surface O atom and one OH$^-$ ion adsorbed on the surface Ti atom. This state 1 is the initial state of the water splitting reaction. From state 1 to state 2, one H$^+$ ion is extracted away from the surface,
$$\rm H_2O/TiO_2 + h^+ \to OH/TiO_2 + H^+. $$
(Adsorption of H$_2$O or OH on the surface we denoted with H$_2$O/TiO$_2$ and OH/TiO$_2$. The hole is denoted by $\rm h^+$.)  In step $2 \to 3$ the second H$^+$ ion is extracted. Next, another water molecule absorbs on the surface (state 4), and the third H$^+$ ion is extracted away from the surface (state 5). From state $5 \to 8$ the fourth H$^+$ ion is extracted away from the surface and one O$_2$ molecule is generated during the process. State 9 is the final state of water splitting reaction without any adsorption. State 9 will go back to state 1 after chemical adsorption of one water molecule.  States 1 through 9 shown in Fig.~\ref{fig:path} are equivalent to states 9 through 17 in Ref.~\onlinecite{doi:10.1021/ja105340b}. The Gibbs energy difference between the beginning state 1 and state 9 is the reaction energy ($\Delta$G) for process
$$2\rm H_2O + 4h^+ \to 4H^+ + O_2.$$
The Gibbs energy difference between state 9 and the last state 1 is the chemisorption energy of one H$_2$O molecule on the surface. 

As already found in Ref.~\onlinecite{doi:10.1021/ja105340b} the rate-controlling step on the pristine surface is the first proton removal step $1 \to 2$.  However, this barrier does not exist on the modified surface, as $1 \to 2$ on the modified surface is exothermic by $-0.48$~eV. Therefore, we find that the modified (001) surface has much better water splitting performance than the pristine surface.

In Table~\ref{table:comparison} we compare barriers computed with and without +{\it U} correction.  As can be seen from the table, the rate-controlling step barrier ($1 \to 2$) is nearly the same with and without the inclusion of +{\it U} correction.

\subsection{Origin of the reduced energy barrier}

\begin{figure}
\includegraphics[width=3.4in]{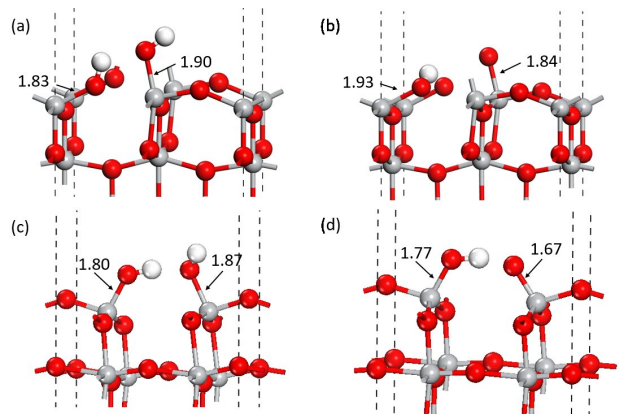}
\caption{\label{fig:structure}Optimized structures of the intermediate states in the reaction path: (a) pristine state 1; (b) pristine state 2; (c) modified state 1; (d) modified state 2. (The bond lengths are given in \AA.)}
\end{figure}

Now we will discuss possible origin of the reduced energy barriers on the modified TiO$_2$ surface. Figure~\ref{fig:structure} shows the optimized structures of states 1 and 2 in the reaction path.  Comparing Figs.~\ref{fig:structure} (b) and (d), we find that the Ti--O bond lengths in state 2 are shorted on the modified state 2 compared to the pristine surface.  We can rationalize this by noting that the surface Ti atom is less saturated on the modified than on the pristine surface.   Therefore, the Ti atoms bind more firmly to O atom and OH$^-$.  As a result, this relatively lower energy of the modified state 2 shown in Fig.~\ref{fig:path} contributes to the reduced energy barrier on the modified TiO$_2$ surface.

Furthermore, as shown in Fig.~\ref{fig:doswater}, the VBM of modified TiO$_2$ surface after adsorbing one H$_2$O is higher than that of pristine TiO$_2$ surface, which is well known to contribute to the improved chemical activity on the surface.\cite{10.2138/am-2000-0416}

\begin{table}
\caption{\label{table:comparison}Comparison of GGA+{\it U} and GGA calculated Gibbs free energy profiles.  Barriers are specified in eV.}
\begin{ruledtabular}
\begin{tabular}{cdddd}
\multirow{2}{*}{Step} & \multicolumn{2}{c}{Pristine surface} & \multicolumn{2}{c}{Modified surface} \\
& \multicolumn{1}{c}{GGA+{\it U}} & \multicolumn{1}{c}{GGA} & \multicolumn{1}{c}{GGA+{\it U}} & \multicolumn{1}{c}{GGA} \\
\hline
$1 \to 2$ & $ 0.44$ & $ 0.55$ & $-0.48$ & $-0.49$\\
$2 \to 3$ & $-1.21$ & $-1.32$ & $-0.41$ & $-0.59$\\
$3 \to 4$ & $-0.20$ & $-0.04$ & $-0.16$ & $-0.11$\\
$4 \to 5$ & $-0.73$ & $-0.42$ & $-0.64$ & $-0.52$\\
$5 \to 6$ & $ 0.25$ & $-0.02$ & $-0.38$ & $ 0.01$\\
$6 \to 7$ & $-0.36$ & $-0.24$ & $-0.01$ & $-0.02$\\
$7 \to 8$ & $-0.48$ & $-0.09$ & $-0.25$ & $ 0.12$\\
$8 \to 9$ & $-0.57$ & $-0.31$ & $-0.56$ & $-1.43$\\
$9 \to 1$ & $-0.39$ & $-0.90$ & $-0.35$ & $-0.14$\\
\end{tabular}
\end{ruledtabular}
\end{table}

\section{\label{sec:conclusion}Outlook and conclusion}

Our calculations show that the modified (001) surface of TiO$_2$ anatase has electronic structure that is favorable for water photocatalysis.  There are several ways to synthesize TiO$_{2}$ (001) films in experiment. One is by adding hydrofluoric acid into the TiO$_2$ precursor, since the hydrofluoric acid can act as a shape controlling agent and makes (001) energetically preferable to (101) facet.\cite{yang2008anatase,doi:10.1021/ja808790p,doi:10.1021/cm200597m,C1CC13929K}     Another is to grow the (001)-oriented anatase TiO$_{2}$ (001) film on a seed layer substrate, such as RbLaNb$_{2}$O$_{7}$,\cite{Nakajima_2011}  Ca$_{2}$Nb$_{3}$O$_{10}$,\cite{doi:10.1021/cg1006204} and amine functionalized glasses.\cite{doi:10.1021/ja0655993}

According to our calculations, the (001) surface of TiO$_2$ anatase can be modified when exposed to the partial pressure of hydrogen atoms about 0.8--80~Pa. Such pressures can be obtained in the experiments. For example, Nakamura\cite{NAKAMURA2000205} synthesized the plasma-treated TiO$_2$ photocatalyst. In his experiment, the chamber pressure was about 270~Pa.   Dobele\cite{Abdel_Rahman_2006}  measured the H$_2$ dissociation degree in processing plasma and found the dissociation ratio is about 5\%. According to this H$_2$ dissociation ratio, the H atom partial pressure is roughly 13~Pa in Nakamura's experiment. Therefore, the hydrogen atom pressure range of 0.8--80~Pa is within the experimental range of achievable hydrogen plasma conditions.

\acknowledgements{This work was supported by Grant No. NSF DMR-1848074.  Computations were performed using the HPCC computer cluster at UCR.}

\bibliography{pap}

\end{document}